# Automated Detection and Diagnosis of Diabetic Retinopathy: A Comprehensive Survey

Vasudevan Lakshminarayanan[1*], Hoda Kheradfallah[1], Arya Sarkar[2], J. Jothi Balaji[3]

[1]Theoretical and Experimental Epistemology Lab, School of Optometry and Vision Science, University of Waterloo, Waterloo, Ontario N2L 3G1, Canada

[2]Department of Computer Engineering, University of Engineering and Management, Kolkata - 700 156, India

[3]Department of Optometry, Medical Research Foundation, Chennai - 600 006, India

*Correspondence: vengulak@uwaterloo.ca

**ABSTRACT**

Diabetic Retinopathy (DR) is a leading cause of vision loss in the world. In the past few years, Artificial intelligence (AI) based approaches have been used to detect and grade DR. Early detection enables appropriate treatment and thus prevents vision loss. Both fundus and optical coherence tomography (OCT) images are used to image the retina. With deep learning / machine-learning-based approaches is possible to extract features from the images and detect the presence of DR. Multiple strategies are implemented to detect and grade the presence of DR using classification, segmentation, and hybrid techniques. This review covers the literature dealing with AI approaches to DR that has been published in the open literature in five-years (2016-2021). In addition, a comprehensive list of available DR datasets is reported. Both the PICO (P-Patient, I-Intervention, C-Control, O-Outcome) and Preferred Reporting Items for Systematic Review and Meta-analysis (PRISMA) 2009 search strategies were employed. We summarize a total of 114 published articles which conformed to the scope of the review. In addition, a list of 43 major datasets is presented.

## 1. Introduction

Diabetic retinopathy (DR) is a major cause of irreversible visual impairment and blindness worldwide [1]. This etiology of DR is due to chronic high blood glucose levels which cause retinal capillary damage. DR mainly affects the working-age population and has a high prevalence rate globally, from 2.6 million in 2015 to 191 million by 2030 [2]. There are two main reasons why early detection is important, DR is difficult to detect in its early stages since there are no visual symptoms. Progression of the disease can lead to blindness. So early diagnosis and regular screening can decrease the risk of visual loss to 57.0 % as well as the cost of treatment [2] DR is clinically graded by observation of the retinal fundus either directly or through imaging techniques such as fundus photography or optical coherence tomography, There are several standard DR grading systems such as the Early Treatment Diabetic Retinopathy Study (ETDRS) grading system [3] with multiple grading levels separating fine detailed DR characteristics evaluated upon all seven retinal fundus fields of view (FOV). Although ETDRS [4] is the gold standard, due to implementation complexity and technical limitations [5], alternative systems are also used such as International Clinical Diabetic Retinopathy (ICDR) [6] scale which is accepted in both clinical and Computer-Aided Diagnosis (CAD) expand settings [7]. The ICDR scale defines 5 severity levels, 4 levels for diabetic macular edema (DME) and requires fewer FOVs [6]. The ICDR levels are discussed below and is illustrated in Fig 1.

- **No Apparent Retinopathy:** No abnormalities.
- **Mild Non-Proliferative Diabetic Retinopathy (NPDR):** This is the first stage of diabetic retinopathy, specifically characterized by tiny areas of swelling in the blood vessels of the retina known as micro aneurysms (MA) [8]. There is an absence of profuse bleeding in the retinal nerves and if DR is detected at this stage, it can help save the patients eyesight with proper medical treatment (Fig 1A)
- **Moderate NPDR:** When left unchecked, mild NPDR progresses is a moderate stage when there is leakage of blood from the blocked retinal vessels. Also, at this stage Hard Exudates (Ex) may exist (Fig 1B). The dilation and constriction of venules in the retina causes venous beading (VB) which are visible ophthalmospically [8].
- **Severe NPDR:** A larger number of blood vessels in the retina are blocked in this stage, causing over 20 intraretinal hemorrhages (IHE; Fig 1C) in all 4 retinal quadrants or there are intraretinal microvascular abnormalities (IRMA) which can be seen as bulges of thin vessels which appear small with and sharp border red spots in at least one quadrant and/or there is definite evidence of VB in over 2 quadrants [8].

- **Proliferative Diabetic Retinopathy (PDR):** This is an advanced stage of the disease and occurs when the condition is left unchecked for an extended period of time. New blood vessels form in the retina and this condition is termed neovascularization (NV). These blood vessels are often fragile, with a consequent risk of fluid leakage and proliferation of fibrous tissue [8]. Different functional visual problems will occur at this stage, such as blurriness, reduced field of vision, and even complete blindness in some cases (Fig D).

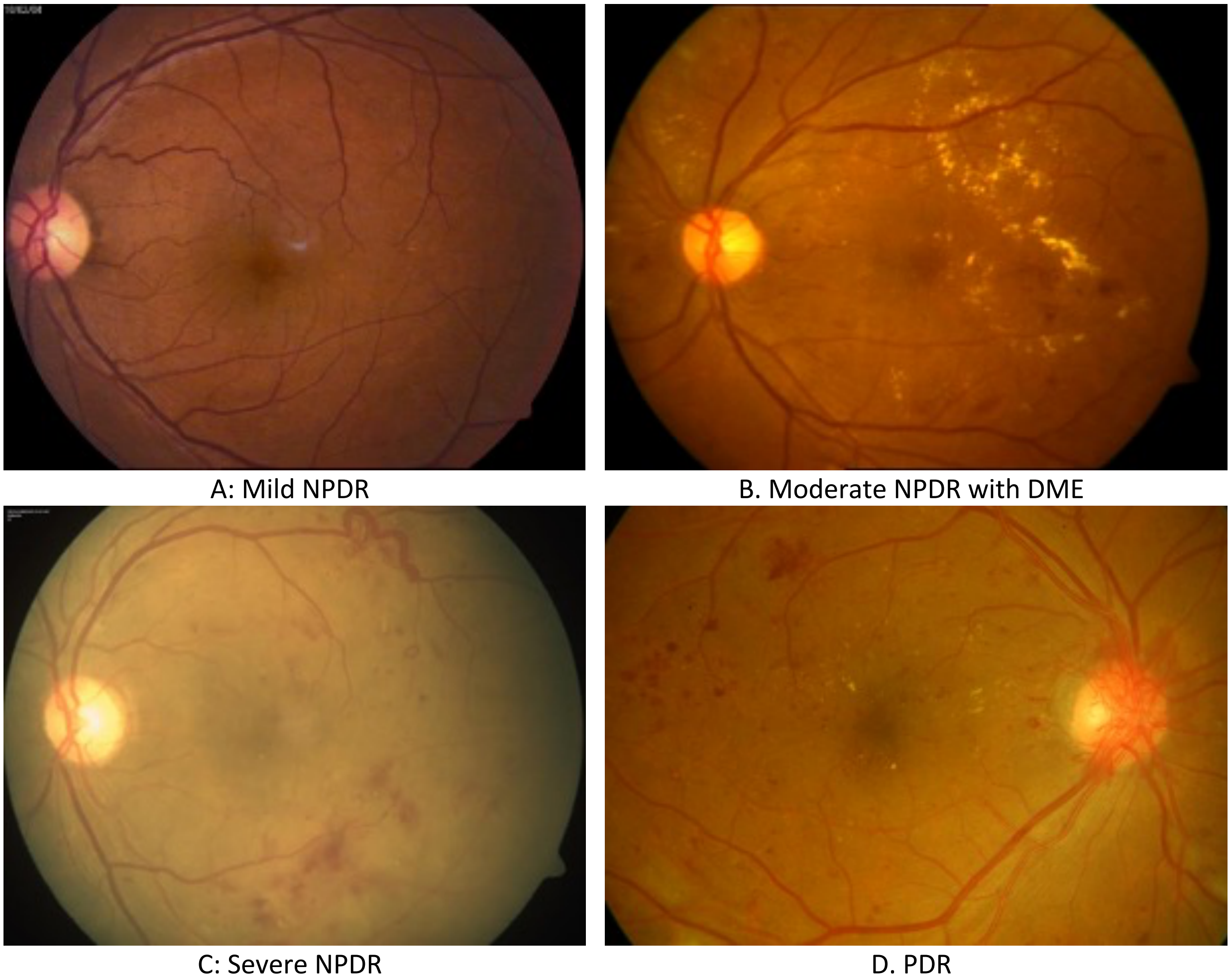

**Fig 1.** Retinal fundus images of different stages of diabetic retinopathy. (A) Stage II: Mild non-proliferative diabetic retinopathy; (B) Stage III: Moderate non-proliferative diabetic retinopathy; (C) Stage IV: Severe non-proliferative diabetic retinopathy; (D) Stage V: Proliferative diabetic retinopathy. (Images curtesy of Rajiv Raman et al. Sankara Nethralaya, India)

The fine pathognomonic DR signs in the initial stages are determined normally, after dilating pupils (mydriasis), DR screening is performed through slit lamp bio-microscopy with a + 90.0 D lens, and direct[9] / indirect ophthalmoscopy to detect [10]. This procedure is time consuming, requires highly trained clinicians who have considerable experience, diagnostic precision that requires time, economic costs and resources. Even if all these are available there is still the possibility of misdiagnosis [11]. This dependency on manual evaluation reading makes the situation more challenging. In year 2020, the number of adults worldwide with DR, and vision-threatening DR was estimated to be 103.12 million, and 28.54 million. By year 2045, the numbers are projected to increase to 160.50 million, and 44.82 million[12]. In addition, in developing countries where there is a shortage of ophthalmologist [13,14] as well as access to standard clinical facilities. This problem also exists in underserved areas of the developed world.

Recent developments in Computer-Aided Diagnosis (CAD) techniques are becoming more prominent in modern ophthalmology [15] as they can save time, cost and human resource for routine DR screening and involve lower diagnostic error factors [15]. CAD can also efficiently manage the increasing number of afflicted DR patients [16] and diagnose DR in early stages with fewer sight threatening effects. These techniques vary depending on the imaging system. The commonly applied imaging methods such as Optical Coherence Tomography (OCT), OCT Angiography (OCTA), Ultrawide-field fundus (UWF) and standard $45^0$ fundus photography are covered in this review. The study performed by Majumder et al[15] reported a smartphone camera screening real time procedure for DR screening.

The main purpose of this review is to analyze 114 articles published within the last 6 years that focus on the detection of DR using CAD techniques. These techniques have made considerable progress in performance with the use of machine learning (ML) and deep learning (DL) schemes that employ the latest developments in deep convolutional neural networks (DCNNs) architectures for DR severity grading, progression analysis, abnormality detection and semantic segmentation. An overview of ophthalmic applications of convolutional neural networks is presented in [17,18,19,20].

## 2. Methods

### 2.1. Literature search details

For this review, literature from 5 publicly accessible databases were surveyed, and these databases were chosen based on their depth, their ease of accessibility, and their popularity. These 5 databases are:

- PubMed: Publications from MEDLINE (https://pubmed.ncbi.nlm.nih.gov/)
- IEEE Xplore: IEEE conference & journals (https://ieeexplore.ieee.org/Xplore/home.jsp)

- PUBLONS: Publications from Web of Science (https://publons.com/about/home/)
- SPIE digital library: Conference & journals from SPIE (https://www.spiedigitallibrary.org/)
- Google Scholar: Database containing conference and journal proceedings from multiple databases (https://scholar.google.co.in/).

Google Scholar has been chosen to fill the gaps in the search strategy by identifying literature from multiple sources, along with articles that might've been missed during manual selection from the other four databases. This review covers the 5-year time-period (2016-2021) so that it is current and the fact that advances in AI enabled diabetic retinopathy detection and grading during this time frame has increased considerably (Fig 2). This figure was generated using the PubMed results.

At the time of writing this review, a total of 10,635 search results were listed in the PubMed database for this time period when just the term *"diabetic retinopathy"* was used. The MEDLINE database is arguably the largest for bio-medical research and are resources from the National Library of Medicine, a part of the U.S. National Institutes of Health.

A search of the IEEE Xplore library and the SPIE digital library for the given time period and search term, report a total of 812 and 332 search results respectively. The IEEE Xplore and SPIE libraries contain only publications of these two professional societies. The other databases further added to this list by listing papers from non-traditional sources such as pre-print servers). In Fig 3 we plot the number of papers published as a function of year using data from all sources.

The scope of this review however is limited to "automated detection and grading of diabetic retinopathy using fundus & OCT images" i.e., the use of artificial intelligence, specifically deep learning, and machine learning techniques to enable detection and grading of diabetic retinopathy in fundus and OCT images. Therefore, to make the search more manageable, a combination of relevant keywords was used using the PICO (P-Patient, I-Intervention, C-Control, O-Outcome) search strategy [21]. The keywords used in the PICO search were predetermined. A combination of ("Diabetic Retinopathy" AND ("deep learning" OR "machine learning" OR "artificial intelligence")) AND (fundus OR OCT) were used which reduced the initial 10,635 search results in PubMed to just 266. during the period under consideration. A manual process of eliminating duplicate search results carried out across the results from all obtained databases resulted in a total number of 114 papers and is represented in the comprehensive.

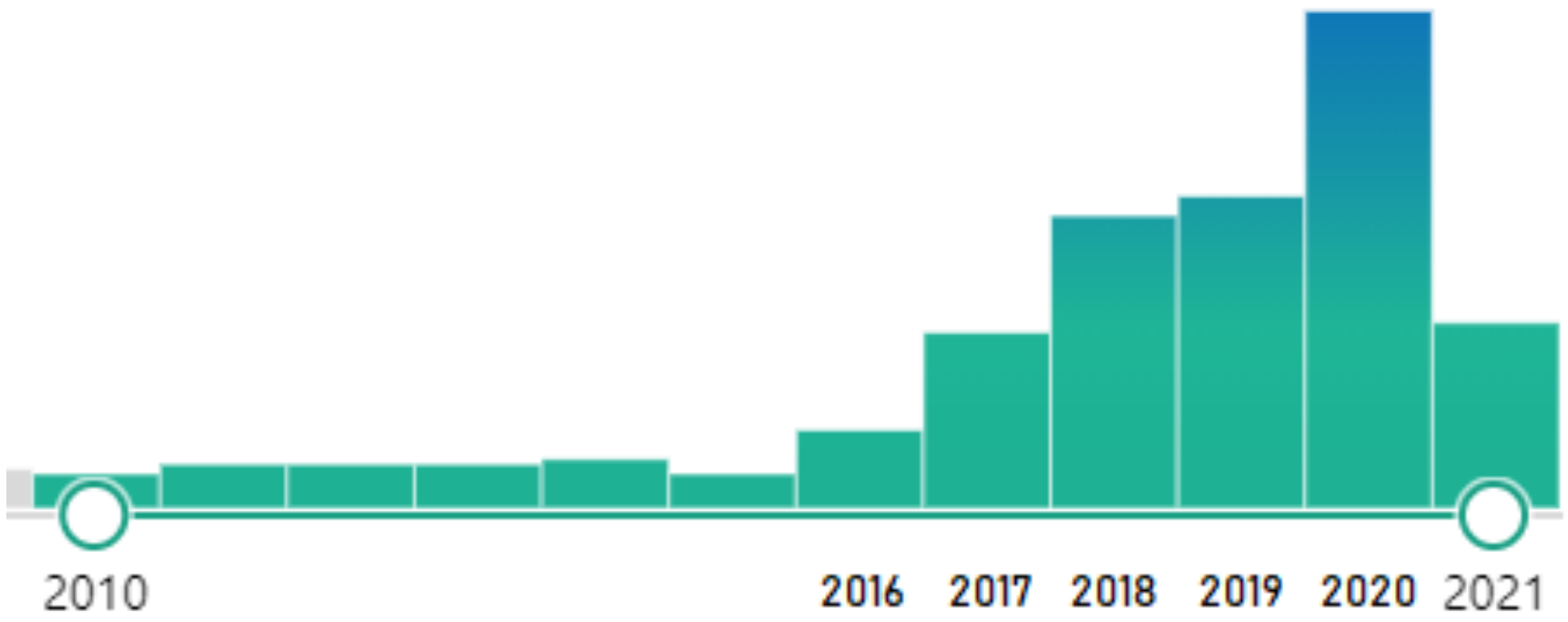

**Fig 2:** Increase in the number of articles matching the predefined keywords over the last 5 years; the PubMed search results were used to create this figure

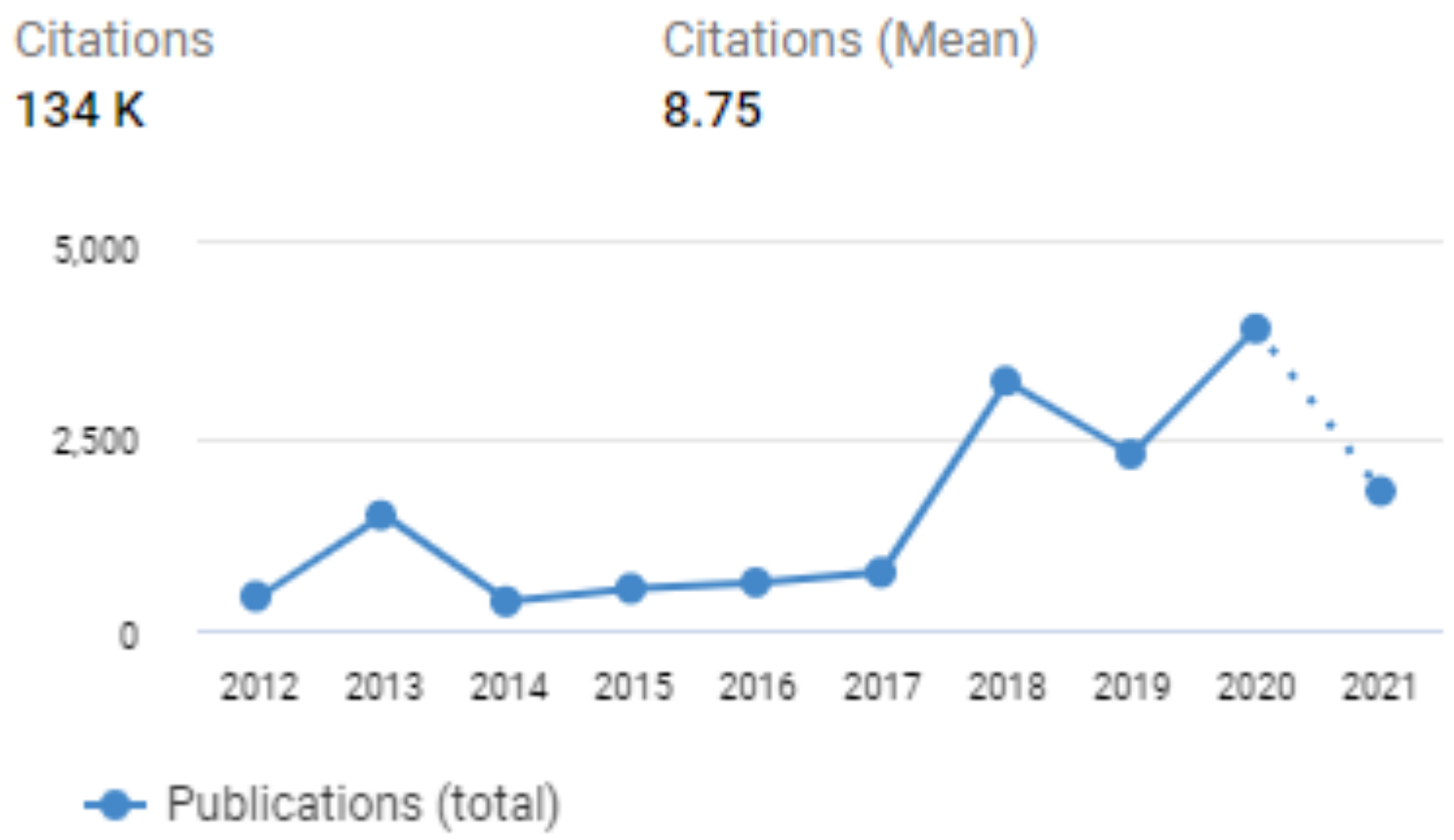

**Fig 3:** A plot of the number of articles as a function of year. This figure was generated using results from all 5 databases using search terms *Diabetic Retinopathy AND (Deep Learning OR Machine Learning)*

The search strategy for identifying relevant research for the review involved three main steps:

1. Using the predefined set of keywords and logical operators, a small set of papers were identified in this time range (2016-21).
2. Using a manual search strategy, the papers falling outside the scope of this review were eliminated.
3. The duplicate articles (i.e. the papers occurring in multiple databases) were eliminated to obtain the set of unique articles used in this review.

   The search strategy followed by this review abides by the Preferred Reporting Items for Systematic Review and Meta-analysis (PRISMA) 2009 checklist [22], and the detailed search and identification pipeline is shown in Fig 4.

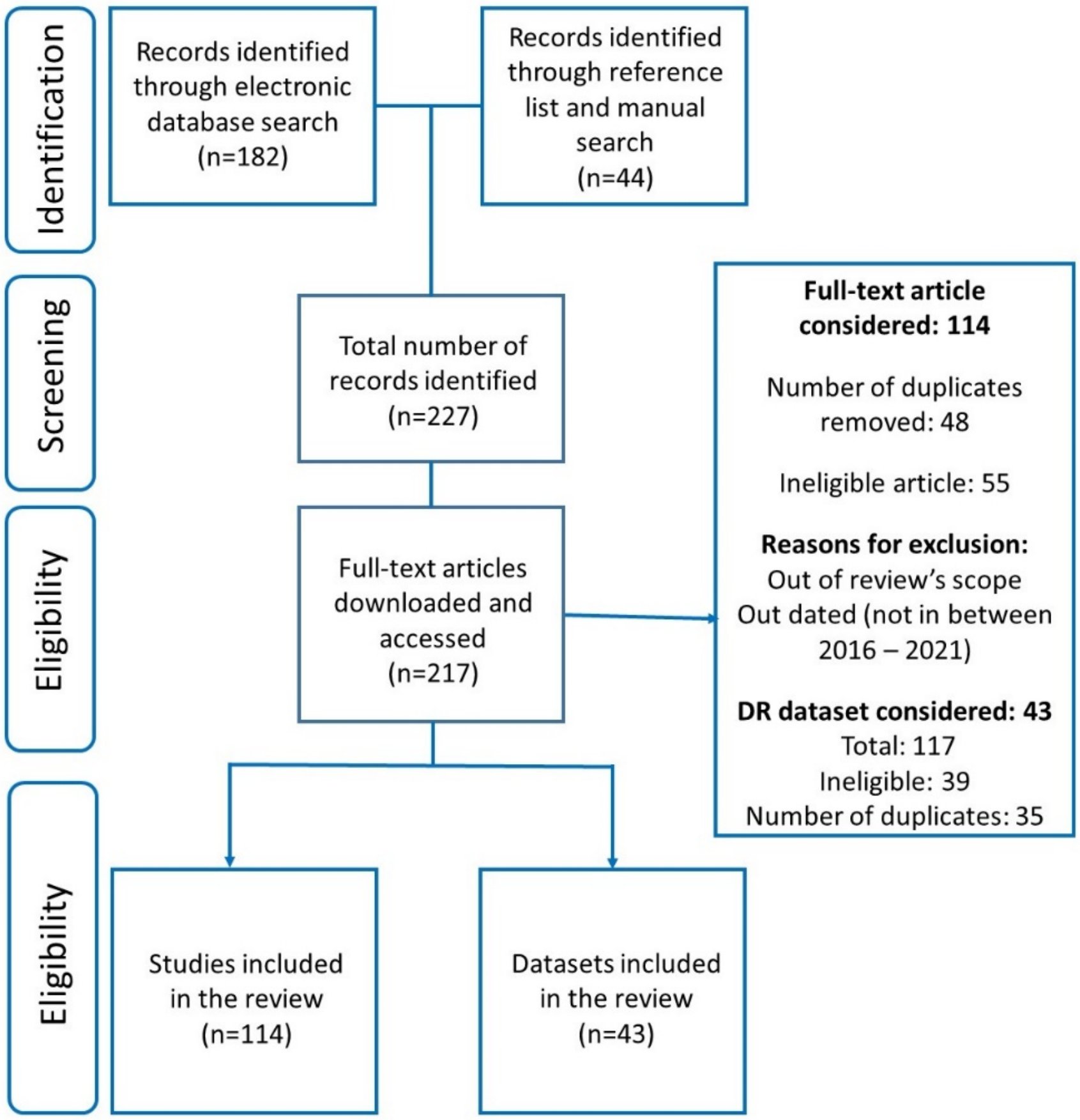

**Fig 4:** Flowchart summarizing the literature search and dataset identification using PRISMA review strategy for identifying articles related to automated detection and diagnosis of diabetic retinopathy.

## 2.2. Dataset search details

The backbone of any automated detection model, be it ML based, DL based, or multi-model-based, is the dataset used for training and validation. High-quality data, which can capture the features of the retina and are correctly graded are of extreme importance in training a good DR detection model. In this review, a comprehensive list of datasets has been created and discussed. A previously published paper[23] also gives a list of ophthalmic image datasets, containing 33 datasets that can be used for training DR detection and grading models. The paper by Khan et al 2021 [23] highlighted 33 of the 43 datasets presented in Table 1. However, some databases which are popular and publicly accessible are not listed by Khan at al 2021 [23] e.g. (UoA-DR [24], Longitudinal DR screening data [25], FGADR [26] etc.), We have identified additional datasets that are available for use.

The search strategy for determining relevant DR detection datasets is as follows:

1. Appropriate results from all 5 of the selected databases (PubMed, PUBLONS, etc.) were searched manually so gather information about names of datasets for DR detection and grading.
2. The original papers and websites associated with each dataset were analyzed and a systematic, tabular representation of all available information was created.
3. The Google dataset search and different forums were checked for missing dataset entries and STEP-2 was repeated for all original datasets found.
4. A final comprehensive list of datasets and its details was generated and represented in Table 1.

A total of 43 datasets were identified employing the search strategy given above. Upon further inspection, a total number of 30 datasets were identified as open access (OA) i.e., can be accessed easily without any permission or payment. Of the total number of datasets, 6 datasets were restricted, however the databases can be accessed with the permission of the author or institution the remaining 7 datasets were private and can't be accessed. These datasets can be used to create a generalized model because of the diversity of images (multi-national, and multi-ethnic groups).

## 3. Results

### 3.1. Dataset search results

This section provides a high-level overview of the search results that were obtained using the datasets as well as, different review articles on datasets in the domain of ophthalmology e.g. Khan et al 2020 [23] and as well as different leads obtained from GitHub and other online forums. In this review, 43 datasets were identified and a general overview of the datasets is systematically presented in this section. The datasets reviewed in this article does not limit to 2016 to 2021 and can be released before that. The list of datasets and their characteristics are shown in Table 1 below. Depending on the restrictions and other proforma required for accessing the datasets, the list has been divided into 3 classes, they are:

- Public open access (OA) datasets with some having high quality DR grading,
- High-quality DR datasets, that can be accessed upon request i.e. can be accessed by filling necessary agreements and forms for fair usage, they are a sub-type of (OA) databases and are termed as access upon request (AUR) in the table.
- Private datasets from different institutions that are not publicly accessible or require explicit permission can access are termed as Not Open Access (NOA).

Table 1: Datasets for DR detection, grading and segmentation and their respective characteristics.

| Dataset | No. of image | Device used | Access | Country | Year | No. Of subjects | Type | Format | Remarks |
|---|---|---|---|---|---|---|---|---|---|
| DRIVE [27] | 40 | Canon CR5 non-mydriatic 3CCD camera with a $45^0$ FOV | OA | Netherlands | 2004 | 400 | Fundus | JPEG | Retinal vessel segmentation and ophthalmic diseases |
| DIARETDB0 [28] | 130 | $50^0$ FOV DFC | OA | Finland | 2006 | NR | Fundus | PNG | DR detection and grading |
| DIARETDB1 [29] | 89 | $50^0$ FOV DFC | OA | Finland | 2007 | NR | Fundus | PNG | DR detection and grading |
| National Taiwan University Hospital[30] | 30 | Heidelberg retina tomography with Rostock corneal module | OA | Japan | 2007-2017 | 30 | Fundus | TIFF | DR, pseudo exfoliation |
| HEI-MED [31] | 169 | Visucam PRO fundus camera (Zeiss, Germany) | OA | USA | 2010 | 910 | Fundus | JPEG | DR detection and grading |
| 19 CF[32] | 60 | NR | OA | Iran | 2012 | 60 | Fundus | JPEG | DR detection |
| FFA Photographs & CF [33] | 120 | NR | OA | Iran | 2012 | 60 | FFA | JPEG | DR grading and lesion detection |
| Fundus Images with Exudates [34] | 35 | NR | OA | Iran | 2012 | NR | Fundus | JPEG | Lesion detection |
| DRiDB [35] | 50 | Zeiss VISUCAM 200 DFC at a 45◦ FOV | AUR | Croatia | 2013 | NR | Fundus | BMP files | DR grading |
| eOphtha [36] | 463 | NR | OA | France | 2013 | NR | Fundus | JPEG | Lesion detection |
| Longitudinal DR screening data [25] | 1120 | Topcon TRC-NW65 with a 45 degrees field of view | OA | Netherlands | 2013 | 70 | Fundus | JPEG | DR grading |
| 22 HRF [37] | 45 | CF-60UVi camera (Canon) | OA | Germany and Czech Republic | 2013 | 45 | Fundus | JPEG | DR detection |
| RITE [38] | 40 | Canon CR5 non-mydriatic 3CCD camera with a $45^0$ FOV | AUR | Netherlands | 2013 | Same As Drive | Fundus | TIFF | Retinal vessel segmentation and ophthalmic diseases |
| DR1[39] | 1077 | TRC-50X mydriatic camera Topcon | OA | Brazil | 2014 | NR | Fundus | TIFF | DR detection |
| DR2 [39] | 520 | TRC-NW8 retinography (Topcon) with a D90 camera (Nikon, Japan) | OA | Brazil | 2014 | NR | Fundus | TIFF | DR detection |
| DRIMDB [40] | 216 | CF-60UVi fundus camera (Canon) | OA | Turkey | 2014 | NR | Fundus | JPEG | DR detection and grading |
| FFA Photographs [41] | 70 | NR | OA | Iran | 2014 | 70 | FFA | JPEG | DR grading and Lesion detection |
| MESSIDOR 1 [42] | 1200 | Topcon TRC NW6 non-mydriatic retinography, $45^0$ FOV | OA | France | 2014 | NR | Fundus | TIFF | DR and DME grading |
| Lotus eyecare hospital [43] | 122 | Canon non-mydriatic Zeiss fundus camera $90^0$ FOV | NOA | India | 2014 | NR | Fundus | JPEG | DR detection |
| Srinivasan [44] | 3231 | SD-OCT (Heidelberg Engineering, Germany) | OA | USA | 2014 | 45 | OCT | TIFF | DR detection and grading, DME, AMD |
| EyePACS [45] | 88,702 | Centervue DRS (Centervue, Italy), Optovue iCam (Optovue, USA), Canon CR1/DGi/CR2 (Canon), and Topcon NW (Topcon) | OA | USA | 2015 | NR | Fundus | JPEG | DR grading |
| Rabbani [46] | 24 images & 24 videos | Heidelberg SPECTRALIS OCT HRA system | OA | USA | 2015 | 24 | OCT | TIFF | Diabetic Eye diseases |
| DR HAGIS[47] | 39 | TRC-NW6s (Topcon), TRC-NW8 (Topcon), or CR-DGi fundus camera (Canon) | OA | UK | 2016 | 38 | Fundus | JPEG | DR, HT, AMD and Glaucoma |
| JICHI DR [48] | 9939 | AFC-230 fundus camera (Nidek) | OA | Japan | 2017 | 2740 | Fundus | JPEG | DR grading |
| Rotterdam Ophthalmic Data Repository DR [49] | 1120 | TRC-NW65 non-mydriatic DFC (Topcon) | OA | Netherlands | 2017 | 70 | Fundus | PNG | DR detection |
| Singapore National DR Screening Program [50] | 4,94,661 | NR | NOA | Singapore | 2017 | 14 880 | Fundus | JPEG | DR, Glaucoma and AMD |
| IDRID [51] | 516 | NR | OA | India | 2018 | NR | Fundus | JPEG | DR grading and lesion segmentation |
| OCTID [52] | 500+ | Cirrus HD-OCT machine (Carl Zeiss Meditec) | OA | Multi ethinic | 2018 | NR | OCT | JPEG | DR, HT, AMD |
| UoA-DR [53] | 200 | Zeiss VISUCAM 500 Fundus Camera FOV $45^0$ | AUR | India | 2018 | NR | Fundus | JPEG | DR grading |
| APTOS [54] | 5590 | DFC | OA | India | 2019 | NR | Fundus | PNG | DR grading |
| CSME [55] | 1445 | NIDEK non-mydriatic AFC-330 auto-fundus camera | NOA | Pakistan | 2019 | NR | Fundus | JPEG | DR grading |
| OCTAGON [56] | 213 | DRI OCT Triton (Topcon) | AUR | Spain | 2019 | 213 | OCTA | JPEG & TIFF | DR detection |
| ODIR-2019 [57] | 8000 | Fundus camera (Canon), Fundus camera (ZEISS), and Fundus camera (Kowa) | OA | China | 2019 | 5000 | Fundus | JPEG | DR, HT, AMD and Glaucoma |
| OIA-DDR [58] | 13,673 | NR | OA | China | 2019 | 9598 | NR | JPEG | DR grading and lesion segmentation |
| Zhongshan Hospital and First People's Hospital [59] | 19,233 | Multiple colour fundus camera | NOA | China | 2019 | 5278 | Fundus | JPEG | DR grading and lesion segmentation |
| AGAR300 [60] | 300 | $45^0$ FOV | OA | India | 2020 | 150 | Fundus | JPEG | DR grading and MA detection |
| Bahawal Victoria Hospital [55] | 2500 | Vision Star, 24.1 Megapixel Nikon D5200 camera | NOA | Pakistan | 2020 | 500 | Fundus | JPEG | DR grading |
| Retinal Lesions [61] | 1593 | Selected from EPACS dataset | AUR | China | 2020 | NR | Fundus | JPEG | DR grading and lesion segmentation |
| Dataset of fundus images for the study of DR [62] | 757 | Visucam 500 camera of the Zeiss brand | OA | Paraguay | 2021 | NR | Fundus | JPEG | DR grading |
| FGADR [58] | 2842 | NR | OA | UAE | 2021 | NR | Fundus | JPEG | DR and DME grading |
| Optos Dataset (Tsukazaki Hospital) [63] | 13047 | 200 Tx ultra-wide-field device (Optos, UK) | NOA | Japan | NR | 5389 | Fundus | JPEG | DR, Glaucoma, AMD, and other eye diseases |
| MESSIDOR 2 [64] | 1748 | Topcon TRC NW6 non-mydriatic retinography $45^0$ FOV | AUR | France | NR | 874 | Fundus | TIFF | DR and DME grading |
| Noor hospital [65] | 4142 | Heidelberg SPECTRALIS SD-OCT imaging system | NOA | Iran | NR | 148 | OCT | TIFF | DR detection |

DFC: Digital fundus camera, RFC: Retinal Fundus Camera, FFA: Fundus Fluorescein Angiogram, DR: Diabetic retinopathy, MA: Microaneurysms, DME: Diabetic Macular Edema, FOV: field-of-view, AMD: Age-related Macular Degeneration; OA: Open access AUR: Access upon request; NOA: Not Open access; CF: Colour Fundus; HT: Hypertension; NR: Not Represented

### 3.2. Diabetic retinopathy classification

This section discusses the classification approaches used for DR detection. The classification can be for the detection of DR [66], referable DR (RDR) [67,68], vision threatening DR (vtDR) [68] or analyze the proliferation level of DR using the ICDR system. Some studies also consider Diabetic Macular Edema (DME) [67,69]. Recent ML and DL methods have produced promising results in automated DR diagnosis.

Thus, multiple performance metrics such as accuracy (ACC), sensitivity (SE) or recall, specificity (SP) or precision, area under the curve (AUC), F1 score and Kappa score are used to evaluate the classification performance. Table 2, 3 present the brief overview on articles that use fundus images for DR classification and articles that classify using novel preprocessing technique. Table 4 lists the recent DR classification studies that use OCT and OCTA images. In the following subsections we provide the details of ML and DL aspects and evaluate the performance of prior studies in terms of quantitative metrics.

### 3.2.1. Machine learning approaches

In this review, 9 out of 93 classification-based studies employed machine learning approaches and 1 article used un-ML method for detecting and grading DR. Hence, in this section, we present the evaluation over various ML based feature extraction and decision making that have been employed in the selected primary studies to construct the DR detection models. In general, six major distinct machine learning algorithms were used in these studies such as feature extraction tools in pre-processing. These are: principal component analysis (PCA) [70,71], linear discriminant analysis (LDA)-based feature selection [71], spatial invariant feature transform (SIFT) [71], support-vector-machine (SVM) [16,71,72,73], KNN [72] and Random Forest (RF) [74]. The methods in combination with DL networks improved the performance and training process. In addition to the widely used ML methods, some studies such as [75] presented a pure ML model with an accuracy of over 80 % including distributed Stochastic Neighbor Embedding (t-SNE) for image dimensionality reduction in combination with ML Bagging Ensemble Classifier (ML-BEC) to improve classification performance using feature bagging technique with low computational time. Ali et al 2020 [55] focused on five fundamental ML models named sequential minimal optimization (SMO), logistic (Lg), multi-layer perceptron (MLP), logistic model tree (LMT), and simple logistic (SLg) in the classification level. This study proposed a novel preprocessing process in which the lesion Region of Interest (ROI) is segmented with a clustering-based method K-Means and extract features of histogram, wavelet, grey scale co-occurrence and run-length matrixes (GLCM and GLRLM) from segmented ROIs. This method outperformed previous models with an average accuracy of 98.83 % with the five ML models. However, a ML model such as SLg performs well, the required classification time is 0.38 with Intel Core i3 1.9 gigahertz (GHz) CPU, 64-bit Windows 10 operating system and 8 gigabytes (GB) memory. This processing time is higher than previous studies. We can also use ML method for OCT and OCTA for DR detection. Recently LiU et al 2021 [76] deployed four ML models of logistic regression (LR), logistic regression regularized with the elastic net penalty (LR-EN), support vector machine (SVM), and the gradient boosting tree named XGBoost with over 246 OCTA wavelet features and obtained ACC, SE and SP of 82 %, 71 % and 77 % respectively. This study, despite inadequate results, has the potential to reach higher scores using model optimization and fine-tuning hyper parameters. These studies show a lower overall performance if using a small number of feature types and simple ML models are used. Dimensionality reduction is an application of ML models which can be added in the decision layer of CAD systems [77,78]. In a support vector machine is used for the classification of features obtained from the state of art DNNs that are optimized with PCA [78]. This provided an accuracy of 85.7 % on preprocessed images. In comparison with methods such as AlexNet, VGG, ResNet, and Inception-v3, the authors report an ACC of 99.5 %. In addition, they also found that this technique is more applicable with considerably less computational cost.

### 3.2.2. Deep learning approaches

This section gives an overview of deep learning algorithms that have been used. Depending on the imaging system used, the image resolution, noise level, contrast as well as the size of the dataset the methods can vary. Some studies propose a customized network such as the work done by Gulshan et al 2016[67], Gargeya et al 2017[66], Rajalakshmi et al 2018[79], Riaz et al 2020[80]. These networks have lower performance outcomes than the state of art networks such as VGG, ResNet, Inception and DenseNet but the fewer layers make them more generalized, suitable for training with small datasets and computationally efficient. Quellec et al 2017 [81] applied L2 regularization over the best performed DCNN in the KAGGLE competition for DR detection named o-O. Another example is that proposed by Sayres et al 2019 [82] which showed 88.4 %, 91.5 %, 94.8 % for ACC, SE and SP respectively over a small subset of 2000 images obtained from the EyePACS database. However, the performance of this network is lower than the results obtained from Mansour et al 2018[71] which used a larger subset of the EyePACS images (35,126 images). Mansour et al 2018[71] also deployed more complex architectures such as the AlexNet on the extracted features of LDA and PCA that generated better results than Sayres et al 2019 [82] with 97.93 %, 100 % and 93 % ACC, SE and SP respectively. Such DCNNs should be used with large datasets since the large number of images used in the training reduces errors. In a deep architecture is applied for a small number of observations it might cause overfitting in which the performance over the test data is not as well as expected on the training data. On the other hand, the deepness of networks does not always guarantee higher performance meaning that they might face with the problems such as vanishing or exploding gradient which will have to be addressed by redesigning the network to simpler architectures. Furthermore, the deep networks extract several low and high-level features. As these image features get more complicated it becomes more difficult to interpret. Sometimes these high-level attributes are not clinically meaningful. For instance, the high-level attributes may refer to an existing bias in all images belonging to a certain class such as light intensity and similar vessel patterns that is not considered as a sign of DR but the DCNN will treat it as a critical feature. Consequently, this fact makes the output prediction erroneous.

In the scope of DL-based classification [83] designed a DL model named Trilogy of Skip-connection Deep Networks (Tri-SDN) over the pretrained base model ResNet50 that applies skip connection blocks to make the tuning faster yielding to ACC and SP of 90.6 % and 82.1 % respectively, which is considerably higher than the values of 83.3 % and 64.1 % compared with skip connection blocks are not used.

There are additional studies that do not focus on proposing new network architecture but enhance the preprocessing output. The study done by Pao et al 2020[84] presents bi-channel customized CNN in which the enhanced image using an image enhancement technique known as unsharp mask and with entropy images used as the inputs of a CNN with 4 convolutional layers with results of 87.83 %, 77.81 %, 93.88 % over ACC, SE and SP. These results are all higher than the case of analysis without preprocessing (81.80 % 68.36 %, 89.87 % respectively).

Shankar et al 2020[85] proposed another approach to preprocessing using Histogram-based segmentation to extract regions containing lesions on fundus images. As the classification step this article utilized Synergic DL (SDL) model and the results indicated that the presented SDL model offers better results than popular DCNNs using MESSIDOR 1 in terms of ACC, SE, SP.

Furthermore, classification is not limited to the DR detection and DCNNs can be applied to detect the presence of DR-related lesions such as that reported by Wang et al 2020 covering twelve lesions: MA, IHE, superficial retinal hemorrhages (SRH), Ex , CWS, venous abnormalities (VAN), IRMA, NV at the disc (NVD), NV elsewhere (NVE), pre-retinal FIP, VPHE, and tractional retinal detachment (TRD) with average precision and AUC 0.67 and 0.95 respectively, however features such as VAN has low individual detection accuracy. This study provides essential steps for DR detection based on the presence of lesions that is more interpretable than DCNNs which act as black boxes [86,87,88].

There are explainable backpropagation-based methods that produce heatmaps of the lesions affecting the classifications DR such as the study done by Keel et al 2019 [89] which highlights Ex, HE and vascular abnormalities in the DR diagnosed images. These methods have limited performance providing generic explanations which might be inadequate to be clinically reliable. Tables 2, 3 and 4 briefly illustrates the previous studies in the scope of DR classification with deep methods.

Table 2: Classification-based studies in DR detection using fundus imaging.

| **Author, Year** | **Dataset** | **Grading details** | **Pre-processing** | **Method** | **Accuracy** | **Sensitivity** | **Specificity** | **AUC** |
|---|---|---|---|---|---|---|---|---|
| Abràmoff , 2016 [68] | MESSIDOR 2 | Detect RDR and vtDR | No | DCNN: IDx-DR X2.1. ML: RF | NA | 96.80% | 87.00% | 0.98 |
| Chandra-kumar, 2016 [90] | EyePACS, DRIVE, STARE | Grade DR based on ICDR scale | Yes | DCNN | STARE and DRIVE:94% | NA | NA | NA |
| Colas, 2016 [91] | EyePACS | Grade DR based on ICDR scale | No | DCNN | NA | 96.20% | 66.60% | 0.94 |
| Gulshan, 2016 [67] | EyePACS, MESSIDOR 2 | Detect DR based on ICDR scale, RDR and referable DME | Yes | DCNN | NA | EyePACS: 97.5% | EyePACS: 93.4% | EyePACS: 0.99 |
| Wong, 2016 [92] | EYEPACS, MESSIDOR 2 | Detect RDR, Referable DME (RDME) | No | DCNN | NA | 90% | 98% | 0.99 |
| Gargeya, 2017 [66] | EyePACS, MESSIDOR 2, eOphtha | Detect DR or non-DR | Yes | DCNN | NA | EyePACS: 94% | EyePACS: 98% | EyePACS: 0.97 |
| So-masundaram, 2017[75] | DIARETDB1 | Detect PDR, NPDR | No | ML: t-SNE and ML-BEC | NA | NA | NA | NA |
| Takahashi, 2017 [48] | Jichi Medical University | Grade DR with the Davis grading scale (NPDR, severe DR,PDR) | No | DCNN: Modified GoogLeNet | 81% | NA | NA | NA |
| Ting, 2017 [30] | SiDRP | Detect RDR, vtDR, glaucoma, AMD | No | DCNN | NA | RDR: 90.5% vtDR: 100% | RDR: 91.6% vtDR: 91.1% | RDR: 0.93 vtDR: 0.95 |
| Quellec, 2017 [81] | EyePACS, eOphta, DIARETDB 1 | Grade DR based on ICDR | Yes | DCNN: L2-regularized o-O DCNN | NA | 94.60% | 77% | 0.955 |
| Wang, 2017 [93] | EyePACS, MESSIDOR 1 | Grade DR based on ICDR scale | Yes | Weakly supervised network to classify image and extract high resolution image patches containing a lesion | MESSIDOR 1: RDR: 91.1% | NA | NA | MESSIDOR 1: RDR:0.957 |
| Benson, 2018 [94] | Vision Quest Biomedical database | Grade DR based on ICDR scale + scars detection | Yes | DCNN: Inception v3 | NA | 90% | 90% | 0.95. |
| Chakrabarty, 2018 [95] | High-Resolution Fundus (HRF) Images | Detect DR | Yes | DCNN | 91.67% | 100% | 100% | F1 score: 1 |
| Costa, 2018 [96] | MESSIDOR 1 | Grade DR based on ICDR scale | No | Multiple Instance Learning (MIL) | NA | NA | NA | 0.9 |
| Dai, 2018 [97] | DIARETDB1 | MA, HE, CWS, Ex detection | Yes | DCNN: Multi-sieving CNN(image to text mapping) | 96.10% | 87.80% | 99.70% | F1 score: 0.93 |
| Dutta, 2018 [98] | EyePACS | Mild NPDR, Moderate NPDR, Severe NPDR, PDR | Yes | DCNN: VGG-Net | 86.30% | NA | NA | NA |
| Kwasigroch, 2018 [99] | EyePACS | Grade DR based on ICDR scale | Yes | DCNN: VGG D | 81.70% | 89.50% | 50.50% | NA |
| Levenkova, 2018 [77] | UWF (Ultra-Wide Field) | Detect CWS, MA, HE, Ex | No | DCNN, SVM | NA | NA | NA | 0.80 |
| Mansour, 2018 [71] | EyePACS | Grade DR based on ICDR scale | Yes | DCNN, ML: AlexNet, LDA, PCA, SVM, SIFT | 97.93% | 100% | 0.93 | NA |
| Rajalakshmi, 2018 [7] | Smartphone-based imaging device | Detect DR and vtDR Grade DR based on ICDR scale | No | DCNN | NA | DR: 95.8% vtDR: 99.1% | DR: 80.2% vtDR: 80.4% | NA |
| Robiul Islam, 2018 [100] | APTOS 2019 | Grade DR based on ICDR scale | Yes | DCNN: VGG16 | 91.32% | NA | NA | NA |
| Zhang, 2018 [101] | EyePACS | Grade DR based on ICDR scale | Yes | DCNN: Resnet-50 | NA | 61% | 84% | 0.83 |
| Zhang, 2018 [102] | EyePACS | Grade DR based on ICDR scale | No | DCNN | 82.10% | 76.10% | 0.855 | Kappa score: 0.66 |
| Arcadu, 2019 [103] | 7 FOV images of RIDE and RISE datasets | 2 step grading based on ETDRS | No | DCNN: Inception v3 | NA | 66% | 77% | 0.68 |
| Bellemo, 2019 [104] | Kitwe Central Hospital, Zambia | Grade DR based on ICDR scale | No | DCNN: Ensemble of Adapted VGGNet & Resenet | NA | RDR: 92·25% vtDR: 99·42% | RDR: 89·04% | RDR: 0·973 vt DR: 0·934 |
| Chowdhury, 2019 [105] | EyePACS | Grade DR based on ICDR scale | Yes | DCNN: Inception v3 | 2 Class: 61.3% | NA | NA | NA |
| Govindaraj, 2019 [106] | MESSIDOR 1 | Detect DR | Yes | Probabilistic Neural Network | 98% | Almost 90% from chart | Almost 97% from chart | F1 score: almost 0.97 |
| Gulshan, 2019 [107] | Aravind Eye Hospital and Sankara Nethralaya, India | Grade DR based on ICDR scale | No | DCNN | NA | Aravind: 88.9% SN: 92.1% | Aravind: 92.2% SN: 95.2% | Quadratic weighted K scores: Aravind: 0.85 SN: 0.91 |
| Hathwar, 2019 [108] | EyePACS, IDRID | Detect DR | Yes | DCNN: Xception-TL | NA | 94.30% | 95.50% | Kappa score: 0.88 |
| He, 2019 [109] | IDRID | Detect DR grade and DME risk | Yes | DCNN: AlexNet | DR grade: 65% | NA | NA | NA |
| Hua, 2019 [83] | Kyung Hee University Medical Center | Grade DR based on ICDR scale | No | DCNN: Tri-SDN | 90.60% | 96.50% | 82.10% | 0.88 |
| Jiang, 2019 [110] | Beijing Tongren Eye Center | DR or Non-DR | Yes | DCNN: Inception v3, Resnet152 and Inception-Resnet- v2 | Integrated model: 88.21% | Integrated model: 85.57% | Integrated model: 90.85% | 0.946 |
| Li, 2019 [111] | IDRID, MESSIDOR 1 | Grade DR based on ICDR scale | No | DCNN: Attention network based on ResNet50 | DR: 92.6%, DME: 91.2% | DR: 92.0% , DME: 70.8% | NA | DR: 0.96 DME: 0.92 |
| Li, 2019 [59] | Shanghai Zhongshan Hospital (SZH) and Shanghai First People's Hospital (SFPH), China, MESSIDOR 2 | Grade DR based on ICDR scale | Yes | DCNN: Inception v3 | 93.49% | 96.93% | 93.45% | 0.9905 |
| Metan, 2019 [112] | EyePACS | Grade DR based on ICDR scale | Yes | DCNN: ResNet | 81% | NA | NA | NA |
| Nagasawa, 2019 [113] | Saneikai Tsukazaki Hospital and Tokushima University Hospital, Japan | Detect PDR | Yes | DCNN: VGG-16 | NA | PDR:94.7% | PDR:97.2% | PDR: 0.96 |
| Qummar, 2019 [114] | EyePACS | Grade DR based on ICDR scale | Yes | DCNN: Ensemble of (Resnet50, Inceptionv3, Xception, Dense121, Dense169) | 80.80% | 51.50% | 86.72% | F1 score: 0.53 |
| Ruamviboonsuk, 2019 [115] | Thailand national DR screening program dataset | Grade DR based on ICDR and detect RDME | No | DCNN | NA | DR: 96.8% | DR: 95.6% | DR: 0.98 |
| Sahlsten, 2019 [69] | Private dataset | Detect DR based on multiple grading systems, RDR and DME | Yes | DCNN: Inception-v3 | NA | 89.60% | 97.40% | 0.98 |
| Sayres, 2019 [82] | EyePACS | Grade DR based on ICDR | No | DCNN | 88.40% | 91.50% | 94.80% | NA |
| Sengupta, 2019 [116] | EyePACS, MESSIDOR 1 | Grade DR based on ICDR scale | Yes | DCNN: Inception-v3 | 90. 4% | 90% | 91.94% | NA |
| Ting, 2019 [117] | SiDRP, SiMES, SINDI, SCES, BES, AFEDS, CUHK, DMP Melb, with 2 FOV | Grade DR based on ICDR scale | Yes | DCNN | NA | NA | NA | Detect DR: 0.86 RDR: 0.96 |
| Zeng, 2019 [118] | EyePACS | Grade DR based on ICDR scale | Yes | DCNN: Inception v3 | NA | 82.2 % | 70.7 % | 0.95 |
| Ali, 2020 [55] | Bahawal Victoria Hospital, Pakistan. | Grade DR based on ICDR scale | Yes | ML: SMO, Lg, MLP, LMT, Lg employed on selected post-optimized hybrid feature datasets | MLP: 73.73%LMT: 73.00 SLg: 73.07 SMO: 68.60 Lg: 72.07% | NA | NA | MLP: 0.916 LMT: 0.919 SLg: 0.921 SMO: 0.878 Lg: 0.923 |
| Araujo, 2020 [119] | EyePACS, MESSIDOR 2, IDRID, DMR, SCREEN-DR, DR1, DRIMDB, HRF | Grade DR based on ICDR scale | Yes | DCNN | NA | NA | NA | Kappa score: EyePAC:0.74 |
| Chetoui, 2020 [24] | EyePACS, MESSIDOR 1, 2, eOphta, UoA-DR from the University of Auckland research, IDRID, STARE, DIARETDB0, 1 | Grade DR based on ICDR scale | Yes | DCNN: Inception-ResNet v2 | 97.90% | 95.80% | 97.10% | 98.60% |
| Elswah, 2020 [73] | IDRID | Grade DR based on ICDR scale | Yes | DCNN: ResNet 50 + NN or SVM | NN: 88% SVM: 65% | NA | NA | NA |
| Gadekallu, 2020 [70] | DR Debrecen dataset collection of 20 features of MESSIDOR 1 | DR or Non-DR | Yes | DCNN ML: PCA+Firefly | 97% | 92% | 95% | NA |
| Gadekallu, 2020 [120] | DR Debrecen dataset | Detect DR | Yes | ML: PCA+ grey wolf optimization (GWO)+ DNN | 97.30% | 91% | 97% | NA |
| Gayathri, 2020 [121] | MESSIDOR 1, EyePACS, DIARETDB0 | Grade DR based on ICDR scale | NA | Wavelet Transform, SVM, RF | MESSIDOR 1: 99.75% | MESSIDOR 1: 99.8% | MESSIDOR 1: 99.9% | NA |
| Jiang, 2020 [122] | MESSIDOR 1 | Image-wise label the presence of MA, HE, Ex, CWS | Yes | DCNN: ResNet 50 based | MA: 89.4% HE: 98.9% Ex: 92.8% CWS: 88.6% Normal: 94.2% | MA: 85.5% HE: 100% Ex: 93.3% CWS: 94.6% Normal: 93.9% | MA: 90.7% HE: 98.6% Ex: 92.7% CWS: 86.8% Normal: 94.4% | MA: 0.94 HE: 1 Ex: 0.97 CWS: 0.97 Normal:0.98 |
| Lands, 2020 [123] | APTOS 2019, APTOS 2015 | Grade DR based on ICDR scale | Ye | DCNN: DensNet 169 | 93% | NA | NA | Kappa score: 0.8 |
| Ludwig, 2020 [10] | EyePACS, APTOS, MESSIDOR 2, EYEGO | Detect RDR | Yes | DCNN: DenseNet201 | NA | MESSIDOR 2: 87% | MESSIDOR 2: 80% | MESSIDOR 2: 0.92 |
| Majumder, 2020 [15] | EyePACS, APTOS 2019 | Grade DR based on ICDR scale | Yes | CNN | 88.50% | NA | NA | NA |
| Memari, 2020 [124] | MESSIDOR 1, HEI-MED | Detect DR | Yes | DCNN | NA | NA | NA | NA |
| Narayanan, 2020 [78] | APTOS 2019 | Detect and grade DR based on ICDR scale | Yes | DCNN: AlexNet,Res Ne, VGG16,Inception v3 | 98.4% | NA | NA | 0.985 |
| Pao, 2020 [84] | EyePACS | Grade DR based on ICDR scale | Yes | CNN: bichannel customized CNN | 87.83% | 77.81% | 93.88% | 0.93 |
| Paradisa, 2020 [72] | DIARETDB 1 | Grade DR based on ICDR scale | Yes | ResNet-50 for extraction and SVM, RF, KNN, and XGBoost as classifiers | SVM: 99%, KNN: 100% | SVM: 99%, KNN: 100% | NA | NA |
| Patel, 2020 [125] | EyePACS | Grade DR based on ICDR scale | Yes | DCNN: MobileNet v2 | 91.29% | NA | NA | NA |
| Riaz, 2020 [80] | EyePACS, MESSIDOR 2 | NA | Yes | DCNN | NA | EyePACS: 94.0% | EyePACS: 97.0% | EyePAC: 0.98 |
| Samanta, 2020 [126] | EyePACS | Grade DR based on ICDR scale | Yes | DCNN: DenseNet121 based | 84.1 % | NA | NA | NA |
| Serener, 2020 [127] | EyePACS, MESSIDOR 1, eOphta, HRF, IDRID | Grade DR based on ICDR scale | Yes | DCNN: ResNet 18 | Country: EyePACS: 65% Continent: EyePACS+HRF:80% | Country: EyePACS: 17% Continent: EyePACS+HRF: 80% | Country: EyePACS: 89% Continent: EyePACS+HRF: 80% | NA |
| Shaban, 2020 [128] | APTOS | Grade DR to non-DR, moderate DR, and severe DR | Yes | DCNN | 88% | 87% | 94% | 0.95 |
| Shankar, 2020 [85] | MESSIDOR 1 | Grade DR based on ICDR scale | Yes | DCNN: Histogram-based segmentation+SDL | 99.28% | 98.54% | 99.38% | NA |
| Singh, 2020 [129] | IDRID, MESSIDOR 1 | Grade DME in 3 levels | Yes | DCNN:Hierarchical Ensemble of CNNs (HE-CNN) | 96.12% | 96.32% | 95.84% | F1 score:0.96 |
| Thota, 2020 [130] | EyePACS | NA | Yes | DCNN: VGG16 | 74% | 80.0 % | 65.0 % | 0.80 |
| Wang, 2020 [131] | 2 Eye hospitals, DIARETDB1, EyePACS, IDRID | MA, HE, EX | Yes | DCNN | MA: 99.7% HE: 98.4% EX: 98.1% Grading: 91.79% | Grading: 80.58% | Grading: 95.77% | Grading:0.93 |
| Wang, 2020 [132] | Shenzhen, Guangdong, China | Grade DR severity based on ICDR scale and detect MA, IHE, SRH, HE, CWS, VAN, IRMA, NVE, NVD, PFP, VPH, TRD | No | DCNN: Multitask network using channel-based attention blocks | NA | NA | NA | Kappa score: Grading:0.80 DR feature: 0.64 |
| Zhang, 2020 [133] | 3 Hospitals in China | Classify to retinal tear & retinal detachment, DR and pathological myopia | Yes | DCNN: InceptionResNetv2 | 93.73% | 91.22% | 96.19% | F1 score: 0.93 |
| Abdelmaksou | EyePACS, MESSIDOR 1, eOphta, CHASEDB 1, | | Yes | U-Net + SVM | 95.10% | 86.10% | 86.80% | 0.91 |
| d , 2021 [134] | HRF, IDRID, STARE, DIARETDB0,1 | | | | | | | |
| Bora, 2021 [115] | EyePACS | Grade DR based on ICDR scale | No | DCNN: Inception v3 | NA | NA | NA | Three FOV: 0·79 One FOV: 0·70 |
| Gangwar, 2021 [135] | APTOS 2019, MESSIDOR 1 | Grade DR based on ICDR scale | Yes | DCNN: Inception ResNet v2 | APTOS:82.18%MESSIDOR 1: 72.33% | NA | NA | NA |
| He, 2021 [136] | DDR, MESSIDOR 1, EyePACS | Grade DR based on ICDR scale | No | DCNN: MobileNet 1 with attention blocks | MESSIDOR 1: 92.1% | MESSIDOR 1: 89.2% | MESSIDOR 1: 91% | F1 score: MESSIDOR 1: 0.89 |
| Hsieh,2021 [30] | National Taiwan University Hospital (NTUH) , Taiwan, EyePACS | Detect any DR, RDR and PDR | Yes | DCNN: Inception v4 for any DR and RDR and ResNet for PDR | Detect DR: 90.7% RDR: 90.0% PDR: 99.1% | Detect DR: 92.2% RDR: 99.2% PDR: 90.9% | Detect DR: 89.5% RDR: 90.1% PDR: 99.3% | 0.955 |
| Khan, 2021[137] | EyePACS | Grade DR based on ICDR scale | Yes | DCNN: customized highly nonlinear scale-invariant network | 85% | 55.6 % | 91.0 % | F1 score: 0.59 |
| Oh , 2021 [2] | 7 FOV fundus images of Catholic Kwandong University, South Korea | Detect DR | Yes | DCNN: ResNet 34 | 83.38% | 83.38% | 83.41% | 0.915 |
| Saeed, 2021 [138] | MESSIDOR, EyePACS | Grade DR based on ICDR scale | No | DCNN: ResNet GB | EyePACS: 99.73% | EyePACS: 96.04% | EyePACS:99.81% | EyePACS: 0.98 |
| Wang, 2021 [139] | EyePACS, images from Peking Union Medical College Hospital, China | Detect RDR with lesion-based segmentation of PHE, VHE, NV, CWS, FIP, IHE, IRMA and MA, then staging based on ICDR scale | No | DCNN: Inception v3 | NA | EyePACS: 90.60% | EyePACS: 80.70% | EyePACS: 0.943 |
| Wang. 2021 [140] | MESSIDOR 1 | Grade DR based on ICDR scale | Yes | DCNN: Multichannel-based GAN with semi super- vision | RDR: 93.2%, DR Grading: 84.23% | RDR: 92.6% | RDR: 91.5% | RDR: 0.96 |

Table 2. Characteristics and evaluation of DR grading methods. In this table, the methods that have no preprocessing or common preprocessing are mentioned. In addition to the abbreviations described earlier this table contains new abbreviations: Singapore integrated Diabetic Retinopathy Screening Program (SiDRP) between 2014 and 2015 (SiDRP 14-15), Singapore Malay Eye Study (SIMES), Singapore Indian Eye Study (SINDI), Singapore Chinese Eye Study (SCES), Beijing Eye Study (BES), African American Eye Study (AFEDS), Chinese University of Hong Kong (CUHK), and Diabetes Management Project Melbourne (DMP Melb) and Generative Adversarial Network (GAN).

Table 3: Classification-based studies in DR detection using a special preprocessing on fundus images for DR grading in ICDR scale.

| Author, Year | Dataset | Pre-processing technique | Method | Accuracy |
|---|---|---|---|---|
| Datta, 2016 [141] | DRIVE, STARE, DIARETDB0, DIARETDB1 | Yes, Contrast optimization | Image processing | NA |
| Lin, 2018 [142] | EyePACS | Yes, Convert to entropy images | DCNN | Original image: 81.8 %<br>Entropy images: 86.1 % |
| Mukhopadhyay, 2018 [143] | Prasad Eye Institute, India | Yes, Local binary patterns | ML: Decision tree, KNN | KNN: 69.8 % |
| Pour, 2020 [144] | MESSIDOR 1,2, IDRID | Yes, CLAHE | DCNN: EfficientNet-B5 | NA |
| Ramchandre, 2020 [145] | APTOS 2019 | Yes, Image augmentation with AUGMIX | DCNN: EfficientNetb3, SEResNeXt32x4d | EfficientNetb3: 91.4 %<br>SEResNeXt32x4d: 85.2 % |
| Shankar, 2020 [85] | MESSIDOR 1 | Yes, CLAHE | DCNN: Hyperparameter Tuning Inception-v4 (HPTI-v4) | 99.5 % |
| Bhardwaj, 2021 [146] | DRIVE, STARE, MESSIDOR 1, DIARETDB1, IDRID, ROC | Yes, Image contrast enhancement and OD localization | DCNN: InceptionResNet v2 | 93.3 % |
| Bilal, 2021 [16] | IDRID | Yes, Adaptive histogram equalization and contrast stretching | ML: SVM + KNN + Binary Tree | 98.1 % |
| Elloumi, 2021 [147] | DIARETDB1 | Yes, Sptic disk location, fundus image partitioning | ML: SVM, RF, KNN | 98.4 % |

AHE: Adaptive Histogram Equalization, CLAHE: Contrast Limited Adaptive Histogram Equalization.

Table 4: Classification-based studies in DR detection using OCT and OCTA.

| Author, Year | Dataset | Grading details | Preprocessing | Method | Accuracy | Sensitivity | Specificity | AUC |
|---|---|---|---|---|---|---|---|---|
| Eladawi, 2018 [148] | OCTA images, University of Louisville, USA | Detect DR | Yes | ML: Vessel segmentation, Local feature extraction, SVM | 97.3% | 97.9% | 96.4% | 0.97 |
| Islam, 2019 [149] | Kermani OCT dataset | NA | Yes | DCNN: DenseNet 201 | 98.6% | 0.986 | 0.995 | NA |
| Le, 2020 [150] | Private OCTA dataset | Grade DR | No | DCNN: VGG16 | 87.3 % | 83.8 % | 90.8 % | 0.97 |
| Sandhu, 2020 [74] | OCT. OCTA, clinical and demographical data, University of Louisville clinical center, USA | Detect mild and moderate DR | Yes | ML: RF | 96.0% | 100.0% | 94.0% | 0.96 |
| Liu, 2021 [76] | Private OCTA dataset | Detect DR | Yes | Logistic Regression (LR), LR regularized with the elastic net (LR-EN), SVM and XGBoost | LR-EN: 80.0% | LR-EN: 82.0% | LR-EN: 84.0% | LR-EN: 0.83 |

### 3.3. Diabetic retinopathy lesion segmentation

The state-of-the-art DR diagnosis machines [66,67,151] dentify referable DR identification without taking lesion information into account. Therefore, their predictions lack clinical interpretation, albeit their high accuracy. This black box nature is the major problem that makes the DCNNs unsuitable for clinical application[86,152,153] and has made the topic of eXplainable AI (XAI) of major importance [153]. Recently, the visualization techniques such as gradient-based XAI have been widely used for evaluating networks. However, these methods with generic heatmaps only highlight the major contributing lesions and hence are not suitable for detection of DR with multiple lesions and severity. Thus, some studies focused on the lesion based DR detection instead. In general, we found 20 papers that do segmentation of the lesions such as MA (10 articles), Ex (9 articles) and IHE, VHE, PHE, IRMA, NV, CWS. In the following sections we discuss the general segmentation approach. The implementation details of each article are accessible in Tables 5 and 6 based on its imaging type.

### 3.3.1. Machine learning and un-machine learning approaches

In general, using ML methods with high processing speed, low computational cost and interpretable decisions are preferred to DCNNs. However, the automatic detection of subtle lesions such as MA did not reach acceptable values. In this review we collected 2 pure ML-involved models and 6 un-ML methods. As reported in a study of Ali Shah 2016 [154] that detects MA with color, Hessian and curvelet-based feature extraction which achieved a SE of 48.2 %. Huang 2018 [155] focused on localizing NV through using Extreme Learning Machine (ELM). This study applied Standard deviation, Gabor, differential invariant and anisotropic filters for this purpose and with the final classifier applying ELM.  This network performed as well as a SVM with lower computational time (6111 s vs 6877 s) with a PC running Microsoft Windows 7 operating system with a Pentium Dual-Core E5500 CPU and 4 GB memory. For the segmentation task, the preprocessing step had a fundamental rule which had a direct effect on the outputs.  The preprocessing techniques varied depending on the lesion type and dataset quality. [Orlando 2018] applied a combination of DCNN and the manually designed features based on illumination correction, CLAHE contrast enhancement and color equalization. Then this high dimensionality feature vector was fed into a RF classifier to detect lesions and achieved an AUC score 0.93 which is comparable with some DCNN models [81,136,140].

Some studies used un-ML methods for detection of exudates such as that by Kaur et al 2018 [156] who proposed a pipeline consisting of vessel and optic disk removal and used a dynamic thresholding method for detection of CWS and Ex. Prior to this study, Imani et al 2016 [157] also did the same process with the focus on Ex with a smaller dataset. In their study they employed additional morphological processing and smooth edge removal to reduce the detection of CWS as Ex. This article reported the SE and SP of 89.1 % and 99.9 % and had almost similar performance compared to Kaur's results with 94.8 %, and 99.8 % for SE and SP respectively. Further description of the recent studies on lesion segmentation with ML approach can be found in Table 5 and 6.

### 3.3.2. Deep learning approaches

Recent works show that DCNNs can produce promising results in automated DR lesion segmentation. DR lesion segmentation is mainly focused on fundus imaging. However, some studies apply a combination of fundus and OCT. Holmberg 2020 [158] proposed a retinal layer extraction pipeline to measure retinal thickness with Unet. Furthermore, Yukun Guo et al 2018[159] applied DCNNs for avascular zone segmentation from OCTA images and received the accuracy of 87.0 % for mild to moderate DR and 76.0 % for severe DR.

Other studies mainly focus on DCNNS applied to fundus images which give a clear view of existing lesions on the surface of retina. Other studies such as Lam et al 2018[160] deployed state of art DCNNS to detect existence of DR lesions in image patches using AlexNet, ResNet, GoogleNet, VGG16 and Inception v3 achieving 98.0 % accuracy on a subset of 243 fundus images obtained from EyePACS. Wang et al 2021[26] also applied Inception v3 as the feature map in combination with FCN 32s as the segmentation part. They reported SE values of 60.7 %, 49.5 %, 28.3 %, 36.3 %, 57.3 %, 8.7 %, 79.8 % and 0.164 over PHE, Ex, VHE, NV, CWS, FIP, IHE, MA respectively. Quellec et al 2017[81] focused on four lesions CWS, Ex, HE and MA using a predefined DCNN architecture named o-O solution and reported the values of 62.4 %, 52.2 %, 44.9 % and 31.6 % over CWS, Ex, HE and MA for SE respectively which shows slightly better performance for CWS and Ex than Wang et al 2021[139]. Considerably better for MA than Wang et al 2021[139]. On the other hand, Wang et al performed better in HE detection. Further details of these article and others can be found in the Tables 5 and 6.

Table 5: Segmentation-based studies in DR detection using fundus images.

| Author, Year | Dataset | Considered lesions | Preprocessing | Segmentation method | Sensitivity/ Specificity | AUC |
|---|---|---|---|---|---|---|
| Imani, 2016 [157] | DIARETDB1, HEI-MED, eOphta | Ex | Yes | Dynamic decision thresholding, morphological feature extraction, smooth edge removal | 89.01%/ 99.93% | 0.961 |
| Shah, 2016 [154] | ROC | MA | Yes | Curvelet transform and rule-based classifier | 48.2 %/ NA | NA |
| Quellec, 2017 [81] | EyePACS, eOphta, DIARETDB1 | CWS, Ex, HE, MA | Yes | DCNN: o-O solution | DIARETDB1:<br>CWS: 62.4%/ NA<br>Ex: 55.2% / NA<br>HE: 44.9% / NA<br>MA: 31.6%/ NA | EyePACS: 0.955 |
| Benzamin, 2018 [161] | IDRID | Ex | Yes | DCNN | 98.29%/ 41.35% | ACC: 96.6% |
| Huang, 2018 [155] | MESSIDOR 1, DIARETDB0,1 | NV | Yes | ELM | NA/ NA | ACC: 89.2% |
| Kaur, 2018 [156] | STARE, eOphta, MESSIDOR 1, DIARETDB1, private dataset | Ex, CWS | Yes | Dynamic decision thresholding | 94.8%/ 99.80% | ACC: 98.43% |
| Lam, 2018 [160] | EyePACS, eOphta | Ex, MA, HE, NV | NA | DCNN: AlexNet, VGG16, GoogLeNet, ResNet, and Inception-v3 | NA/ NA | EyePACS:0.99 ACC:98.0 % |
| Orlando,2018 [162] | eOphtha, DIARETDB1, MESSIDOR 1 | MA, HE | Yes | ML: RF | NA/ NA | 0.93 |
| Eftekhari, 2019 [163] | ROC, eOphta | MA | Yes | DCNN: Two level CNN, thresholded probability map | NA/ NA | ROC: 0.660 |
| Wu, 2019 [164] | HRF | Blood vessels, optic disc and other regions | Yes | DCNN: AlexNet, GoogleNet, Resnet50, VGG19 | NA/ NA | AlexNet: 0.94 ACC: 95.45% |
| Yan, 2019 [165] | IDRID | Ex, MA, HE, CWS | Yes | DCNN: Global and local Unet | NA/ NA | Ex:0.889 MA:0.525 HE: 0.703 CWS:0.679 |
| Qiao, 2020 [166] | IDRID | MA | Yes | DCNN | 98.4%/ 97.10% | ACC: 97.8% |
| Wang, 2021 [140] | EyePACS, images from Peking Union Medical College Hospital | Detect RDR with lesion-based segmentation of PHE, VHE, NV, CWS, FIP, IHE, Ex, MA | No | DCNN: Inception v3 and FCN 32s | PHE: 60.7% / 90.9% Ex: 49.5% / 87.4% VHE: 28.3% / 84.6% NV: 36.3% / 83.7% CWS: 57.3% / 80.1% FIP: 8.7% / 78.0% IHE: 79.8% / 57.7% MA: 16.4%/ 49.8% | NA |
| Wei, 2021 [61] | EyePACS | MA, IHE, VHE, PHE, Ex, CWS, FIP, NV | Yes | DCNN: Transfer learning from Inception v3 | NA/ NA | NA |
| Xu, 2021 [167] | IDRID | Ex, MA, HE, CWS | Yes | DCNN: Enhanced Unet named FFUnet | Ex:87.55%/ NA MA:59.33% / NA HE:73.42% / NA CWS:79.33%/ NA | IOU: Ex:0.84 MA:0.56 HE:0.73 CWS:0.75 |

ICDR scale: International Clinical Diabetic Retinopathy scale, RDR: Referable DR, vtDR: vision threatening DR, PDR: Proliferative DR, NPDR: Non-Proliferative DR, MA: Microaneurysm, Ex: hard Exudate, CWS: Cotton Wool Spot, HE: Hemorrhage, FIP: Fibrous Proliferation, VHE: Vitreous Hemorrhage, PHE: Preretinal Hemorrhage, NV: Neovascularization.

Table 6: Segmentation-based studies in DR detection using OCT, OCTA images.

| **Author, Year** | **Dataset** | **Considered lesions** | **Pre-processing** | **Segmentation method** | **Sensitivity** | **Specificity** | **AUC** |
|---|---|---|---|---|---|---|---|
| Guo, 2018 [159] | UW-OCTA private dataset | Avascular area | Yes | DCNN | Control: 100.0% Diabetes without DR: 99.0% Mild to moderate DR: 99.0% Severe DR: 100.0% | Control: 84.0% Diabetes without DR: 77.0% Mild to moderate DR: 85.0% Severe DR: 68.0% | ACC: Control: 89.0% Diabetes without DR: 79% Mild to moderate DR: 87% Severe DR: 76.0% |
| ElTanboly, 2018 [168] | OCT and OCTA images of University of Lousville | 12 different retinal layers & segmented OCTA plexuses | No | SVM | NA | NA | ACC: 97.0% |
| ElTanboly, 2018 [169] | SD-OCT images of Kentucky Lions Eye Center | 12 distinct retinal layers | Yes | Statistical analysis and extraction of features such as tortuosity, reflectivity, and thickness for 12 retinal layers | NA | NA | ACC:73.2% |
| Sandhu, 2018 [168] | OCT images of University of Louisville, USA | 12 layers; quantifies the reflectivity, curvature, and thickness | Yes | DCNN: 2 Stage deep CNN | 92.5% | 95.0% | ACC: 93.8% |
| Holmberg, 2020 [158] | OCT from Helmholtz Zentrum München, Fundus from EyePACS | Segment retinal thickness map, Grade DR based on ICDR scale | No | DCNN: On OCT: Retinal layer segmentation with Unet On fundus: Self supervised learning, ResNet50 | NA | NA | IOU: on OCT: 0.94 |

## 4. Conclusion

Recent studies for DR detection are mainly focused on automated methods know as CAD systems. In the scope of CAD system for DR there are two major approaches known as first classification and staging DR severity and second segmentation of lesions such as MA, HE, Ex, CWS associated with DR.

The DR databases are categorized to public databases (36 out of 43) and private databases (7 out of 43). These databases contain fundus and OCT retinal images and among these two imaging modalities fundus photos are used in 86.0 % of the published studies. Several public large fundus datasets are available online. The images might have been taken with different systems that affect image quality. Furthermore, some of the image-wise DR labels can be erroneous. The image databases that provide lesion annotations constitute only a small portion of the databases that require considerable resources for pixel-wise annotation. Hence, some of them contain fewer images than image-wise labeled databases. Furthermore, Lesion annotations requires intra-annotator agreement and high annotation precision. These factors make the dataset error sensitive, and its quality evaluation might become complicated.

The DR classification needs a standard grading system validated by clinicians. The ETDRS is the gold standard grading system proposed for DR progression grading but since this grading type needs fine detail evaluation and access to all 7 FOV fundus images, these issues make the use of ETDRS limited. There ICDR with less precise scales is applicable for 1 FOV images to detect the DR severity levels.

The classification and grading DR images can be divided to two main approaches, namely ML-based and DL-based classification. The ML-based DR detection has considerably better performance than grading using the ICDR scale which needs to extract higher-level features associated with each level of DR [55,70]. The evaluation results also proved that the DCNN architectures can achieve higher performance scores when large databases are used [71]. There is a trade-off between the performance on one side and the architecture complexity, processing time and the lack of interpretability over the network's decisions and extracted features on the other side. Thus, some recent works have proposed semi-DCNN models containing both DL based and ML-based models acting as classifier or feature extractor [70,71]. The use of regularization techniques is another solution to reduce the complexity of DCNN models [81].

The second approach for CAD-related studies in DR is pixel-wise lesion segmentation or image-wise lesion detection. The main lesions of DR are MA, Ex, HE, CWS. These lesions have different detection difficulty which directly affects the performance of proposed pipeline. Among these lesions the annotation of MA is more challenging [26,167]. Since this lesion is difficult to detect and is the main sign of DR in early stages, some studies focused on the pixel-wise segmentation of this lesion with DCNNs and achieved high enough scores [166]. Although some of the recent DCNN- based works exhibit high performance in term of the standard metrics, the lack of interpretability may not be sufficiently valid for real-life clinical applications. This interpretability brings into the picture the concept of explainable AI or XAI. Explainability studies aim to show the features that influence the decision of a model the most. Singh et al [170] have reviewed the currently used explainability methods. There is also the need for a large fundus database with high precision annotation of all associated DR lesions to help in designing more robust and pipelines with high performance.

**Author Contributions:** Conceptualization V.L. and J.J.B.; methodology, V.L. and J.J.B.; dataset constitution H.K. and A.S.; writing-original draft preparation H.K, A.S. and J.J.B; writing-review and editing V.L; project administration V. L. and J. J. B. All authors have read and agreed to the published version of the manuscript.

**Funding:** This research was partly supported by a DISCOVERY grant to VL from the Natural Sciences and Engineering Research Council of Canada.

**Institutional Review Board Statement:** Not applicable. Study doesn’t involve humans or animals.

**Informed Consent Statement:** Study didn’t involve humans or animals and therefore informed consent is not applicable.

**Data Availability Statement:** This is review of the published articles, and all the articles are publically available.

**Acknowledgments:** A. S. acknowledges MITACS, Canada for the award of a summer internship.

**Conflicts of Interest:** The authors have no conflicts of interest.